\documentclass[aip, jcp, preprint,showpacs,superscriptaddress,groupedaddress]{revtex4-1}
\usepackage{natbib}
\usepackage{braket}
\usepackage{bm}
\usepackage{bbm}
\usepackage{amsmath}
\usepackage{amssymb}

\begin{document}

\title{Bivariational Principle for an Antisymmetrized Product of Nonorthogonal Geminals Appropriate for Strong Electron Correlation}
\author{Paul A. Johnson}
 \email{paul.johnson@chm.ulaval.ca}
 \affiliation{D\'{e}partement de chimie, Universit\'{e} Laval, Qu\'{e}bec, Qu\'{e}bec, G1V 0A6, Canada}
\author{Paul W. Ayers}
 \email{ayers@mcmaster.ca}
 \affiliation{Department of Chemistry \& Chemical Biology, McMaster University, 1280 Main St. W, Hamilton, Ontario, L8S 4M1, Canada}
\author{Stijn De Baerdemacker}
 \affiliation{Department of Chemistry, University of New Brunswick, Fredericton, New Brunswick, E3B 5A3, Canada}
\author{Peter A. Limacher}
 \affiliation{SAP Security Research, Karlsruhe, Germany}
\author{Dimitri Van Neck}
 \affiliation{Center for Molecular Modeling, Ghent University, Technologiepark 46, 9052 Zwijnaarde, Belgium}

\date{\today}

\begin{abstract}
We develop a bivariational principle for an antisymmetric product of nonorthogonal geminals. Special cases reduce to the antisymmetric product of strongly-orthogonal geminals (APSG), the generalized valence bond-perfect pairing (GVB-PP), and the antisymmetrized geminal power (AGP) wavefunctions. The presented method employs wavefunctions of the same type as Richardson-Gaudin (RG) states, but which are not eigenvectors of a model Hamiltonian which would allow for more freedom in the mean-field. The general idea is to work with the same state in a primal picture in terms of pairs, and in a dual picture in terms of pair-holes. This leads to an asymmetric energy expression which may be optimized bivariationally, and is strictly variational when the two representations are consistent. The general approach may be useful in other contexts, such as for computationally feasible variational coupled-cluster methods.

\end{abstract}

\maketitle

\section{Introduction} \label{sec:OS_1}
Most computational methods for the electronic structure theory of molecules and solids are based
on the orbital (for molecules) or band (for solids) picture \cite{helgaker_book,raghavachari,headgordon,helgaker}. In this picture, the ground-state wavefunction is approximated by a Slater determinant of the single-electron eigenfunctions (orbitals) from a mean-field model (e.g., Hartree-Fock, Kohn-Sham). The electrons are assumed to move independently, with each electron feeling only the average repulsion from the other electrons in the system. In many systems this is a reasonable first approximation and a single Slater determinant wavefunction is a good starting point to develop the wavefunction. The orbital picture is particularly reliable for equilibrium thermodynamic properties of organic molecules but is much less accurate for chemical transition states, electronic excited states, and inorganic substances containing $d$-block and $f$-block elements.

Strongly-correlated systems require \emph{many} Slater determinants for even a qualitatively correct physical description.\cite{garnet2,anisimov} This is a clear indication that Slater determinants are not an efficient basis and the mean-field behaviour is not independent electrons. Conventional density-based and wavefunction-based methods are usually unreliable for strong correlation, and the methods that are appropriate are usually computationally expensive. Large molecules and complex
materials with thousands of valence electrons can be routinely modelled with the orbital picture, but no such tools exist when the orbital picture fails. Our ultimate goal is to develop new models for strongly correlated systems with hundreds of valence electrons.

Because modelling strongly correlated substances is difficult, condensed-matter physicists usually model these systems by introducing model Hamiltonians that capture the key qualitative features of the system in question. We have recently shown how these model Hamiltonians can also be used to provide quantitative predictions for real physical systems \cite{johnson,limacher,kasia_hubbard,pawel_geminal}. The basic idea is to (1) find a model Hamiltonian that reproduces the key qualitative features of the system of interest, (2) use the wavefunction-form of that model Hamiltonian to model the substance.

In our previous work\cite{johnson:2020,fecteau:2020,fecteau:2021,johnson:2021,moisset:2022,fecteau:2022} we have chosen wavefunctions that are eigenvectors of a model Hamiltonian,
\begin{align} \label{eq:OS_1}
\hat{H}_{model}(\bm{\eta})\ket{\Psi_{model}(\bm{\eta})} &= E_{model}\ket{\Psi_{model}(\bm{\eta})}
\end{align}
and minimized the expectation value of our target Hamiltonian with respect to the parameters, $\bm{\eta}$, in the model Hamiltonian,
\begin{align} \label{eq:OS_2}
E_{GS} \approx \min_{\bm{\eta}} \braket{\Psi_{model}(\bm{\eta}) | \hat{H} | \Psi_{model}(\bm{\eta}) }.
\end{align}
Because these model Hamiltonians are exactly solved by the algebraic Bethe Ansatz (ABA) \cite{korepin_book}, we refer to the eigenfunctions of the model Hamiltonian as on-shell Bethe vectors. A variational method based on on-shell Bethe vectors has been implemented previously and shown to be quite accurate provided the correct on-shell vector is chosen. The relevant details are summarized briefly in section \ref{sec:OS_3}.

The requirement that the wavefunction be an eigenfunction of the model Hamiltonian may be relaxed\cite{johnson,limacher,moisset:2022} yielding \emph{off-shell} Bethe vectors. Variational optimization of off-shell Bethe vectors is much more difficult (and perhaps often impossible), so our previous work used the projected Schr\"{o}dinger equation to establish a system of nonlinear
equations that can be solved for the parameters in the off-shell Bethe vectors,
\begin{align} \label{eq:OS_3}
\braket{\Phi | \hat{H} | \Psi(\bm{\eta})} = E\braket{\Phi | \Psi(\bm{\eta})}.
\end{align}

The primary goal of this paper is to present a bivariational principle for off-shell Bethe vectors; this is presented in section \ref{sec:OS_4}. The basic idea is to rewrite the variational principle as,
\begin{align} \label{eq:OS_4}
E_{GS} = \min_{\substack{\braket{\tilde{\Psi}(\bm{\eta}) | \Psi(\bm{\eta}) }=1 \\ \ket{\Psi(\bm{\eta})}=\ket{\tilde{\Psi}(\bm{\eta})}  } }
\braket{\tilde{\Psi}(\bm{\eta}) | \hat{H} | \Psi(\bm{\eta}) }.
\end{align}
That is to say, we take two states $\ket{\Psi(\bm{\eta})}$ and $\ket{\tilde{\Psi}(\bm{\eta})}$, and minimize \eqref{eq:OS_4} while attempting to ensure that the two states are the same. This will be explained in section \ref{sec:OS_7p2}.

This approach is applicable to any ABA solvable model Hamiltonian, though we focus for the moment on the Richardson-Gaudin family of Hamiltonians \cite{gaudin2,rich1,rich2,rich3,dukelsky3}. The necessary background material is presented in section \ref{sec:OS_2}. After discussing the on-shell and off-shell formulations, we discuss the extension to nonorthogonal orbitals in section \ref{sec:OS_6}. The bivariational principle for off-shell RG states and a different approach to the antisymmetrized geminal power wavefunction are presented in sections \ref{sec:OS_7} and \ref{sec:OS_8}, respectively. Notes on the computational algorithm are presented in section \ref{sec:OS_9}, and we discuss possible generalizations in section \ref{sec:OS_10}.

\section{Background} \label{sec:OS_2}
The systems we wish to solve are described by the Coulomb Hamiltonian,
\begin{align} \label{eq:OS_5}
\hat{H}_C = \sum_{ij} h_{ij} \sum_{\sigma} a^{\dagger}_{i\sigma} a_{j\sigma} +\frac{1}{2} \sum_{ijkl}V_{ijkl} \sum_{\sigma \tau} a^{\dagger}_{i\sigma} a^{\dagger}_{j\tau} a_{l\tau} a_{k\sigma},
\end{align}
where $a^{\dagger}_{i\sigma}$($a_{i\sigma}$) are the operators for creating (destroying) an electron in spatial orbital $i$ with spin projection $\sigma$. In Eq. \eqref{eq:OS_5}, the integrals are expressed in physicists' notation
\begin{align} \label{eq:OS_6}
h_{ij} &= \int d\mathbf{r} \phi^*_i (\mathbf{r}) \left( - \frac{1}{2} \nabla^2 - \sum_I \frac{Z_I}{| \mathbf{r} - \mathbf{R}_I |} \right) \phi_j (\mathbf{r}) \\
V_{ijkl} &= \int d\mathbf{r}_1 d\mathbf{r}_2 \frac{\phi^*_i(\mathbf{r}_1)  \phi^*_j(\mathbf{r}_2)  \phi_k(\mathbf{r}_1)  \phi_l(\mathbf{r}_2)  }{| \mathbf{r}_1 - \mathbf{r}_2|}.
\end{align}
With a wavefunction ansatz $\ket{\Psi}$ the energy is computed,
\begin{align} \label{eq:OS_7}
E[\Psi] = \frac{ \braket{\Psi | \hat{H}_C | \Psi}} {\braket{\Psi|\Psi}} 
= \sum_{ij}h_{ij} \sum_{\sigma}\gamma^{\sigma}_{ij}[\Psi] +\frac{1}{2} \sum_{ijkl}V_{ijkl} \sum_{\sigma\tau}\Gamma^{\sigma\tau}_{ijkl}[\Psi]
\end{align}
in terms of the normalized 1- and 2-electron reduced density matrices (RDMs),
\begin{align} \label{eq:OS_9}
\gamma^{\sigma}_{ij}[\Psi] &= \braket{\Psi | a^{\dagger}_{i\sigma} a_{j\sigma} | \Psi} \\
\Gamma^{\sigma\tau}_{ijkl} [\Psi]  &= \braket{\Psi | a^{\dagger}_{i\sigma}a^{\dagger}_{j\tau} a_{l\tau} a_{k\sigma} | \Psi}.
\end{align}
Because it is extremely difficult to determine, much less optimize, $E[\Psi]$ for arbitrary
wavefunction forms, most practical approaches to the quantum many-body problem simplify the form of either the Hamiltonian or the wavefunction. We do not want to surrender our ability to accurately model real physical systems, so we will only approximate the wavefunction
in this paper.

The particular model wavefunctions we consider are built from fully-paired states: they are geminal products. \cite{hurley,mcweeny3,parr,parks1,surjan4,surjan7,silver4,mehler,silver5,kutz2,miller1,coleman1,coleman2}. It is convenient to work with objects that create/destroy pairs:
\begin{align} \label{eq:OS_11}
S^+_i = a^{\dagger}_{i\uparrow}a^{\dagger}_{i\downarrow},\quad S^-_i = a_{i\downarrow}a_{i\uparrow},\quad S^z_i = \frac{1}{2} \left(a^{\dagger}_{i\uparrow}a_{i\uparrow} +a^{\dagger}_{i\downarrow}a_{i\downarrow} -1 \right)
\end{align}
with the structure,
\begin{align} \label{eq:OS_12}
[S^z_i,S^{\pm}_j] &= \pm \delta_{ij} S^{\pm}_i, \quad
[S^+_i,S^-_j] = 2\delta_{ij} S^z_i. 
\end{align}
The pairing scheme may be engineered to be more general \cite{bajdich2,limacher,scuseria1}) though the algebraic structure of the states does not change. The operators
$S^{\pm}_i$ create/destroy a pair in state $i$, while $S^z_i$ counts the number of pairs in state $i$. It is convenient to work with the number operator
\begin{align}
\hat{n}_i = 2 S^z_i + 1.
\end{align}

The most general geminal mean-field wavefunction is the antisymmetric product of interacting geminals (APIG)\cite{silver2,silver1} for $N_P$ pairs in $N_{orb}$ spatial orbitals,
\begin{align} \label{eq:OS_13}
\ket{\text{APIG}} = \prod^{N_P}_{\alpha=1}\sum^{N_{orb}}_{i=1} g^i_{\alpha}S^+_i\ket{\theta}. 
\end{align}
The state $\ket{\theta}$ represents the vacuum; it ordinarily represents the physical vacuum but it is only required to be a vacuum with respect to the removal of electron
pairs. For APIG, the amplitudes $g^i_{\alpha}$ have no additional structure. Equation \eqref{eq:OS_13} can be expanded as a linear combination of $\binom{N_{orb}}{N_P}$ Slater determinants
\begin{align} \label{eq:OS_14}
\ket{\text{APIG}} = \sum_{\substack{ m_i =\{0,1\} \\ \sum^{N_{orb}}_{i=1} m_i =N_P   }} 
\left\vert \mathbf{C}(\mathbf{m})  \right\vert^+ 
\left( S^+_1 \right)^{m_1} \left( S^+_2 \right)^{m_2} \dots \left( S^+_{N_{orb}} \right)^{m_{N_{orb}}} \ket{\theta}.
\end{align}
Each expansion coefficient is the permanent of an
$N_P \times N_P$ matrix consisting of the geminal
amplitudes of the occupied (i.e., $m_i =1$) orbital pairs,
\begin{align} \label{eq:OS_15}
\left\vert \mathbf{C}(\mathbf{m})  \right\vert^+  =
\left\vert \begin{array}{cccc}
g^{i_1}_{1} & g^{i_2}_{1} & \dots & g^{i_{N_P}}_{1} \\
g^{i_1}_{2} & g^{i_2}_{2} & \dots & g^{i_{N_P}}_{2} \\
\vdots & & \ddots & \vdots \\
g^{i_1}_{N_P} & g^{i_2}_{N_P} & \dots & g^{i_{N_P}}_{N_P}
\end{array}  \right\vert^+  \\
1=m_{i_1}=m_{i_2}=\dots=m_{i_{N_P}}. \nonumber
\end{align}
Evaluating the permanent of a matrix is \#P-hard \cite{valiant}, so the APIG wavefunction is computationally intractable. For certain special forms of $g^i_{\alpha}$, however, the permanents may be efficiently evaluated \cite{johnson,limacher,borchardt,carlitz}.

It has long been recognized that for antisymmetric products of geminals (Eq. \eqref{eq:OS_13}) \cite{kutz1} and
doubly-occupied configuration interaction wavefunctions (Eq. \eqref{eq:OS_14}) \cite{weinhold,weinhold:1967b,cook} have especially simple
formulas for the reduced density matrices. Specifically, the energy expression becomes
\begin{align}
E = 2 \sum_i h_{ii} \gamma_i + \sum_{i \neq j} (2V_{ijij} - V_{ijji}) D_{ij} + \sum_{ij} V_{iijj} P_{ij}
\end{align}
in terms of the normalized RDM elements
\begin{subequations} \label{eq:OS_16}
\begin{align} 
\gamma_i &= \frac{1}{2} \frac{\braket{\Psi | \hat{n}_i | \Psi}}{\braket{\Psi| \Psi}} \\
D_{ij} &= \frac{1}{4} \frac{\braket{\Psi | \hat{n}_i \hat{n}_j | \Psi}}{\braket{\Psi|\Psi}} \\
P_{ij} &= \frac{\braket{\Psi|S^+_i S^-_j | \Psi}}{\braket{\Psi|\Psi}}.
\end{align}
\end{subequations}
The 1-RDM $\gamma_i$ is diagonal, while the 2-RDM has two non-zero pieces: the \emph{diagonal correlation function} $D_{ij}$ and the \emph{pair correlation function} $P_{ij}$. Notice that the diagonal terms refer to the same matrix element $P_{ii} = D_{ii}$ so to avoid double counting this term is assigned to $P_{ii}$. Elements of the 1-RDM represent the probability of occupying the $i$th orbital. The elements $D_{ij}$ represent the probability of occupying the $i$th and $j$th orbitals simultaneously, while elements of $P_{ij}$ represent the probability of transferring a pair from level $i$ to level $j$.

\section{RG states: On-Shell Bethe Vectors of the reduced BCS Hamiltonian (Cauchy Geminals)} \label{sec:OS_3}
Since we are interested in systems that are well-described by pairing wavefunctions \eqref{eq:OS_13}, we need Hamiltonians whose eigenstates are of this form. One such Hamiltonian is the reduced Bardeen-Cooper-Schrieffer (BCS) Hamiltonian \cite{bardeen:1957a,BCS,gaudin2,rich1,rich2,rich3},
\begin{align} \label{eq:OS_22}
\hat{H}_{BCS} = \frac{1}{2} \sum^{N_{orb}}_{i=1} \varepsilon_i \hat{n}_i - \frac{g}{2} \sum_{ij}S^+_iS^-_j.
\end{align}
The reduced BCS Hamiltonian is believed to provide qualitative insights into strongly correlated systems like superconducting nanograins \cite{dukelsky3,dukelsky4,kruchinin,vonDelft}. Richardson\cite{rich1,rich2,rich3} showed that the eigenvectors of the reduced BCS Hamiltonian are products of geminals:
\begin{align} \label{eq:OS_23}
\ket{\{ u \}} = \prod^{N_P}_{a=1} S^+(u_a)\ket{\theta}
\end{align}
where the operators $S^+(u_a)$ create a pair delocalized over the entire single particle space, i.e.
\begin{align} \label{eq:OS_24}
S^+(u_a) = \sum^{N_{orb}}_{i=1} \frac{S^+_i}{u_a-\varepsilon_i}.
\end{align}
The complex numbers $\{u_a \}^{N_P}_{a=1}$ are called by various names, including rapidities, quasimomenta, spectral parameters, and pairing energies. The parameters in the reduced BCS Hamiltonian, $\{\varepsilon_{i} \}^{N_{orb}}_{i=1}$ and $g$, represent single-particle energies and the pairing strength respectively (cf. Eq. \eqref{eq:OS_22}). The two-electron states generated by Eq. \eqref{eq:OS_24} are called Cauchy geminals because $1/(u_a-\varepsilon_i)$ are the elements of a Cauchy matrix. The form of the solution, Eq. \eqref{eq:OS_23}-\eqref{eq:OS_24}, is also referred to as the rational solution for the XXX (isotropic) Richardson-Gaudin model.

The wavefunction form \eqref{eq:OS_23} is a particular example of the ABA: acting with the Hamiltonian \eqref{eq:OS_22} on the state \eqref{eq:OS_23} and collecting terms yields 
\begin{align}
\hat{H}_{BCS} \ket{\{u\}} = \sum_a u_a \ket{\{u\}} - \frac{g}{2} \sum_i S^+_i \sum_a \Lambda(u_a) \prod_{b \neq a} S^+(v_b) \ket{\theta}
\end{align}
one term proportional to \eqref{eq:OS_23} and $N_P$ linearly independent terms which vanish provided that
\begin{align} \label{eq:OS_25}
\Lambda(u_a) &= \frac{2}{g} + \sum^{N_{orb}}_{i=1}\frac{1}{u_a -\varepsilon_i} + \sum_{b\neq a} \frac{2}{u_b -u_a} =0.
\end{align}
The set of equations \eqref{eq:OS_25} are Richardson's equations and must be solved numerically. Many algorithms are available.\cite{rombouts,faribault:2011,faribault_DG,stijn_PRC,guan:2012,pogosov:2012,claeys:2015}

Computationally practical expressions for the RDM elements have been derived previously, so we will
only present the main ideas. The engine driving all of these results is Slavnov's theorem \cite{slavnov,zhou,claeys:2017b,belliard:2019}, which gives a formula for the inner product of two RG states, Eqs. \eqref{eq:OS_23}-\eqref{eq:OS_24}, where one of the sets of rapidities (here $\{u\}$) is a solution of Richardson's equations \eqref{eq:OS_25}, with the other set arbitrary. Specifically,
\begin{align} \label{eq:OS_26}
\braket{\{u\} | \{v\}} &= \frac{\prod_{a , b} \left(u_{a}-v_{b} \right)}{\prod_{a<b} \left(v_{a}-v_{b} \right)  \left(u_{b} - u_{a} \right)}\det J\left(\{u\},\{v\} \right)
\end{align}
where the matrix $\mathbf{J}$ has elements
\begin{align} \label{eq:OS_27}
J_{ab} = \frac{1}{u_{a}-v_{b}}\left(\sum_{i}\frac{1}{\left(u_{a}-\varepsilon_{i}\right)\left(v_{b}-\varepsilon_{i}\right)} -2\sum_{c \neq a} \frac{1}{\left(u_{a}-u_{c}\right)\left(v_{b}-u_{c}\right)} \right).
\end{align}
Choosing $\{v\} =\{u\}$ gives the norm of an RG state,
\begin{align} \label{eq:OS_28}
\braket{\{u\} | \{u\} } &= \det G
\end{align}
where the elements of the matrix G are
\begin{align} \label{eq:OS_29}
G_{ab} &=
\begin{cases}
 \sum^{N}_{i=1} \frac{1}{(u_{a}-\varepsilon_{i})^{2}} -2\sum^{N_{P}}_{c\neq a}\frac{1}{(u_{a}-u_{c})^{2}} & a=b\\ \frac{2}{(u_{a}-u_{b})^{2}} & a\neq b
\end{cases}.
\end{align}
This expression for the norm was originally derived by Richardson by an alternative method \cite{rich3}. The \emph{Gaudin matrix}, $G$, is the Jacobian of Richardson's equations. I.e.,
\begin{align} \label{eq:OS_30}
G_{ab} &= \frac{\partial \Lambda(u_a)}{\partial u_b}
\end{align}
where $\Lambda(u_a)$ is one of Richardson’s  equations, cf. Eq. \eqref{eq:OS_25}. The expressions for the RDM elements are corollaries to Slavnov's theorem obtained from the form factor approach. For the 1-RDM, we can use the commutation relations \eqref{eq:OS_12} to move $\hat{n}_i$ to the right until it acts on the vacuum, obtaining
\begin{align} \label{eq:OS_31}
\frac{1}{2}\braket{\{u\} | \hat{n}_i | \{v\}} &=
\sum_a\frac{\braket{\{u\} | S^+_i | \{v\}_a}}{v_a -\varepsilon_i}.
\end{align}
Here $\ket{\{v\}_a}$ denotes the $(N_P-1)$-pair state with the rapidity $v_a$ removed, and the scalar product $\braket{\{u\}|S^+_i|\{v\}_a}$ is called a form factor. Similarly, the 2-RDM elements are obtained from
\begin{align} \label{eq:OS_32}
\frac{1}{4}\braket{\{u\} | \hat{n}_i \hat{n}_j | \{v\}} &= 
\sum_{a\neq b}\frac{\braket{\{u\} | S^+_iS^+_j | \{v\}_{a,b}} }{(v_a -\varepsilon_i)(v_b-\varepsilon_j)}
\end{align}
and
\begin{align} \label{eq:OS_33}
\braket{\{u\} | S^+_iS^-_j | \{v\}} &= 
\sum_a\frac{\braket{\{u\} | S^+_i | \{v\}_a}}{v_a -\varepsilon_j} 
- \sum_{a\neq b}\frac{\braket{\{u\} | S^+_iS^+_j | \{v\}_{a,b}}}{(v_a -\varepsilon_j)(v_b-\varepsilon_j)}
\end{align}
where $\{v\}_{a,b}$ is the set $\{v\}$ without $v_a$ and $v_b$. The expression \eqref{eq:OS_33} appears asymmetric in $i$ and $j$ since all the terms arise from moving $S^-_j$ to the right, but from many numerical tests\cite{faribault1,faribault2,johnson:2020,moisset:2022} it is clear that it is in fact symmetric as it should be.

The geminal-creation operators, Eq. \eqref{eq:OS_24}, have simple poles whenever the rapidity coincides with a single-particle energy $\{\varepsilon\}$. The residues of the poles are the pair creators
\begin{align} \label{eq:OS_36}
S^+_i &= \lim_{v\rightarrow\varepsilon_i} (v-\varepsilon_i) S^+(v).
\end{align}
The form factors that appear in Eqs. \eqref{eq:OS_31}-\eqref{eq:OS_33} can therefore be evaluated as the corresponding residues of the scalar product, $\braket{\{u\}|\{v\}}$
\begin{align} \label{eq:OS_37}
\braket{\{u\} | S^+_i | \{v\}_a} = \lim_{v_a\rightarrow\varepsilon_i}(v_a-\varepsilon_i)\braket{\{u\} | \{v\}}
\end{align}
\begin{align} \label{eq:OS_38}
\braket{\{u\} | S^+_iS^+_j | \{v\}_{a,b}} = \lim_{v_a\rightarrow\varepsilon_i}\lim_{v_b\rightarrow\varepsilon_j}(v_a-\varepsilon_i)(v_b-\varepsilon_j)\braket{\{u\}|\{v\}}.
\end{align}
Expressions for $\gamma_i$, $D_{ij}$, and $P_{ij}$ are directly obtained by making the substitution $\{v\}\rightarrow\{u\}$ and normalizing. The form factors become ratios of determinants differing by one (or two) columns
\begin{align}
\frac{\braket{\{u\} | S^+_i      | \{u\}_a}}{\braket{\{u\}|\{u\}}} &= \frac{\det G^i_a}{\det G}\\
\frac{\braket{\{u\} | S^+_i S^+_j| \{u\}_{a,b}}}{\braket{\{u\}|\{u\}}} &= \frac{\det G^{ij}_{ab}}{\det G}.
\end{align}
The matrix $G^i_a$ is the matrix in \eqref{eq:OS_29} with the $a$th column replaced by the $i$th vector
\begin{align} \label{eq:rhs}
\textbf{r}_i = 
\begin{pmatrix}
\frac{1}{(u_1 - \varepsilon_i)^2} \\
\frac{1}{(u_2 - \varepsilon_i)^2} \\
\vdots \\
\frac{1}{(u_{N_P} - \varepsilon_i)^2} \\
\end{pmatrix}
\end{align}
while $G^{ij}_{ab}$ is \eqref{eq:OS_29} with the $a$th column replaced by the $i$th version of \eqref{eq:rhs} and the $b$th column replaced by the $j$th version of \eqref{eq:rhs}. It is not difficult to verify that the double column replacement can be simplified to a $2 \times 2$ determinant of single column replacements,
\begin{align}
\frac{\det G^{ij}_{ab}}{\det G} = \frac{\det G^i_a}{\det G} \frac{\det G^j_b}{\det G} - \frac{\det G^i_b}{\det G}\frac{\det G^j_a}{\det G}.
\end{align}
and rather than compute the $N_P \times N_{orb}$ determinants $G^i_a$, we can solve the $N_{orb}$ sets of linear equations
\begin{align}
G \textbf{x} = \textbf{r}_i.
\end{align}
From Cramer's rule, the solutions of these linear equations directly give the ratios of determinants
\begin{align}
\textbf{x}_a = \frac{\det G^i_a}{\det G}.
\end{align}
The RDMs for on-shell RG states are thus constructed from solutions of linear equations. No determinants are computed numerically.

\section{Dual Construction} 
\subsection{Rational Off-Shell Bethe Vectors} \label{sec:OS_4}

In the previous section we considered wavefunctions that are eigenfunctions of the
reduced BCS Hamiltonian, for which there were $N_{orb}-1$ free
parameters: the zero of energy and energy scale for the
reduced BCS Hamiltonian are arbitrary. A more flexible wavefunction can be obtained by not
requiring that the rapidities satisfy Richardson's equations. Wavefunctions of this form are \emph{off-shell} Bethe vectors, and there are $N_P + N_{orb}-1$ free parameters.

Although Slavnov's theorem for the scalar product only applies when at least one of the
states is an on-shell Bethe vector, the expressions \eqref{eq:OS_31}-\eqref{eq:OS_33} and \eqref{eq:OS_36}-\eqref{eq:OS_38} remain valid for off-shell Bethe vectors \cite{faribault3}. Therefore, in order to evaluate the energy of an off-shell Bethe vector, we
only require an expression for the scalar product $\braket{\{u\} | \{v\}}$. One could attempt to evaluate this
scalar product by expanding the states $\ket{\{u\}}$ and $\ket{\{v\}}$ in Slater determinants (cf. Eq. \eqref{eq:OS_14}), but this is impractical because (a) there are a \emph{large} number of terms in the sum and (b)
evaluating the permanent is \#P hard. The second problem is circumvented because
$\mathbf{C}(\mathbf{m})$ is a Cauchy matrix: Borchardt's theorem establishes that the permanent of a Cauchy matrix can be computed as the ratio of determinants \cite{borchardt},
\begin{align} \label{eq:OS_39}
\vert\mathbf{C}(\mathbf{m})\vert^+ = \frac{\vert\mathbf{C}(\mathbf{m})* \mathbf{C}(\mathbf{m})\vert}{\vert\mathbf{C}(\mathbf{m})\vert}
=\vert \mathbf{C}^{-1}(\mathbf{m}) \cdot (\mathbf{C}(\mathbf{m}) * \mathbf{C}(\mathbf{m}))\vert
\end{align}
where $\mathbf{A} * \mathbf{B} $ denotes the Hadamard (i.e., elementwise) product of two matrices and $\mathbf{A}\cdot\mathbf{B}$ denotes the conventional matrix multiplication. However, we still need to find a
method to avoid summing over the factorial number of terms in the Slater determinant
expansion.

To this end, we \emph{assume} that there exists a dual representation for the state of interest: it can be written either as the creation of
$N_P$ pairs on a physical vacuum,
\begin{align} \label{eq:OS_40}
\ket{\{v\}} = \prod^{N_P}_{a=1} S^+(v_{a})\ket{\theta}
\end{align}
or as the annihilation of $N_{orb}-N_P$ pairs from the pseudovacuum, $\ket{\tilde{\theta}}$, in which every orbital is
doubly-occupied,
\begin{align} \label{eq:OS_41}
\ket{\{v\}} = \prod^{N_{orb}-N_P}_{a=1} S^-(\tilde{u}^*_a)\ket{\tilde{\theta}},
\end{align}
in terms of a set of dual rapidities $\{\tilde{u}\}$ where
\begin{align} \label{eq:OS_42}
S^-(\tilde{u}^*_a)=\sum^{N_{orb}}_{i=1} \frac{S^-_i}{\tilde{u}^*_a-\varepsilon_i} = (S^+(\tilde{u}_a))^{\dagger}.
\end{align}
The scalar product can therefore be written as
\begin{align} \label{eq:OS_43}
\braket{\{\tilde{u}\} | \{v\}} = \braket{\tilde{\theta} | \prod^{N_{orb}-N_P}_{a=1} S^+(\tilde{u}_a) \prod^{N_P}_{b=1}S^+(v_b) |\theta}
\end{align}
which is the projection of a state with $N_{orb}$ pairs onto the Slater determinant $\ket{\tilde{\theta}}=\ket{1\bar{1}2\bar{2}\dots N_{orb}\bar{N}_{orb}}$. The coefficient in Eq. \eqref{eq:OS_14} corresponds to a vector $\mathbf{m}$ in which every
element is 1, and $\vert\mathbf{C}(\mathbf{1})\vert^+$ can be evaluated in $\mathcal{O}(N^3_{orb})$ cost using Eq. \eqref{eq:OS_39}. Faribault and Schuricht\cite{faribault3} (and Gaudin\cite{gaudin_book2}) worked out an explicit formula for this permanent
\begin{align} \label{eq:OS_44}
\braket{\{\tilde{u}\} | \{v\}} = \det \mathbf{K}
\end{align}
where the elements of the matrix $\mathbf{K}$ are given by
\begin{align} \label{eq:OS_45}
k_{ab} = 
\begin{cases}
\sum_{k\neq a} \frac{1}{\varepsilon_a -\varepsilon_k} 
- \sum^{N_P}_{\beta=1}\frac{1}{\varepsilon_a -v_{\beta}} 
- \sum^{N_{orb}-N_P}_{\beta=1}\frac{1}{\varepsilon_a -\tilde{u}_{\beta}} & a=b \\
\frac{1}{\varepsilon_a-\varepsilon_b} & a \neq b
\end{cases}.
\end{align}
If $\bra{\{\tilde{u}\}}=\bra{\{v\}}$, then Eq. \eqref{eq:OS_44} is a formula for the scalar product,
$\braket{\{v\} | \{v\}}=\braket{\{\tilde{u}\} | \{v\}}$. It is important that the single-particle energies, $\{\varepsilon\}$, be the same in both $S^+(\{v\})$ and
$S^-(\{\tilde{u}^*\})$; if this were not true, then a factorial number of permanents would need to be evaluated to
compute the norm, and any hope of favourable scaling would be lost. 

Whether $\bra{\{\tilde{u}\}}=\bra{\{v\}}$ or not, the expressions for the form factors are obtained by
methods very similar to those in the previous section. Just like Eq. \eqref{eq:OS_37}, one has
\begin{align} \label{eq:OS_46}
\braket{\{\tilde{u}\} | S^+_i | \{v\}_{\alpha}} = \lim_{v_{\alpha}\rightarrow\varepsilon_i}(v_{\alpha}-\varepsilon_i)\det\mathbf{K}
=\det \mathbf{K}_{i\alpha}
\end{align}
where $\mathbf{K}_{i\alpha}$ is the $N_{orb}-1 \times N_{orb}-1$ matrix with elements
\begin{align} \label{eq:OS_47}
[k_{i\alpha}]_{ab} =
\begin{cases}
\sum_{k\neq i,a} \frac{1}{\varepsilon_a -\varepsilon_k} 
- \sum_{\gamma\neq \alpha}\frac{1}{\varepsilon_a -v_{\gamma}} 
- \sum^{N_{orb}-N_P}_{\beta=1}\frac{1}{\varepsilon_a -\tilde{u}_{\beta}} & a=b \neq i \\
\frac{1}{\varepsilon_a-\varepsilon_b} & a \neq b (\neq i)
\end{cases}
\end{align}
To derive Eq. \eqref{eq:OS_47}, one Laplace-expands the determinant from Eq. \eqref{eq:OS_46} along the $i$th row. Then, taking the residue $v_{\alpha}\rightarrow \varepsilon_i$, the only non-vanishing cofactor is proportional to $k_{ii}$ . The other diagonal elements become $(j\neq i)$
\begin{align} \label{eq:OS_48}
\lim_{\lambda_{\alpha}\rightarrow\varepsilon_i} k_{jj} = 
\sum_{k\neq j} \frac{1}{\varepsilon_j -\varepsilon_k} 
- \sum_{\gamma\neq \alpha}\frac{1}{\varepsilon_a -\lambda_{\gamma}} 
- \sum^{N_{orb}-N_P}_{\beta=1}\frac{1}{\varepsilon_a -\tilde{\mu}_{\beta}} -\frac{1}{\varepsilon_j-\varepsilon_i}
\end{align}
and since
\begin{align} \label{eq:OS_49}
\lim_{\lambda_{\alpha}\rightarrow\varepsilon_i}(\lambda_{\alpha}-\varepsilon_i) k_{ii} =1,
\end{align}
one obtains Eq. \eqref{eq:OS_47}. Analogous to Eq. \eqref{eq:OS_38} one has,
\begin{align} \label{eq:OS_50}
\braket{\{\tilde{u}\} | S^+_iS^+_j | \{v\}_{\alpha,\beta}} = \lim_{v_{\alpha}\rightarrow\varepsilon_i}\lim_{v_{\beta}\rightarrow\varepsilon_j}(v_a-\varepsilon_i)(v_b-\varepsilon_j)\braket{\{\mu\} | \{\lambda\}}
= \det \mathbf{K}_{i\alpha,j\beta},
\end{align}
where $\mathbf{K}_{i\alpha,j\beta}$ is the $N_{orb}-2 \times N_{orb}-2$ matrix with elements
\begin{align} \label{eq:OS_51}
[k_{i\alpha,j\beta}]_{ab} =
\begin{cases}
\sum_{k\neq i,j,a} \frac{1}{\varepsilon_a -\varepsilon_k} 
- \sum_{\gamma\neq \alpha,\beta}\frac{1}{\varepsilon_a -v_{\gamma}} 
- \sum^{N_{orb}-N_P}_{\beta=1}\frac{1}{\varepsilon_a -\tilde{u}_{\beta}} & a=b \neq i,j \\
\frac{1}{\varepsilon_a-\varepsilon_b} & a \neq b (\neq i,j)
\end{cases}
\end{align}
Whereas for on-shell RG states, the matrices required to compute RDM elements collapsed together naturally leading to linear equations, here, there is certainly structure, but it doesn't condense as naturally. The residues are co-factors of the common matrix $\mathbf{K}$ with an additional diagonal update. As a diagonal update is rank $N_{orb}$ rather than rank 1, it is not clear how the ratios of determinants could simplify. In any case the matrix elements are expressible as linear combinations of ratios of determinants. Again, if the states are dual, the matrix elements are RDM elements. Otherwise they represent transition density matrix elements. In particular,
\begin{align} \label{eq:OS_52}
\gamma^{\tilde{u},v}_i := \frac{1}{2} \frac{\braket{\{\tilde{u}\} | \hat{n}_i | \{v\}}}{\braket{\{\tilde{u}\}|\{v\}}} &= \sum^{N_P}_{\alpha=1} \frac{1}{(v_{\alpha}-\varepsilon_i)}\frac{\det \mathbf{K}_{i\alpha}}{\det\mathbf{K}} \\
D^{\tilde{u},v}_{ij} := \frac{1}{4} \frac{\braket{\{\tilde{u}\} | \hat{n}_i \hat{n}_j | \{v\}}}{\braket{\{\tilde{u}\}|\{v\}}} 
&= \sum^{N_P}_{\alpha=1}\sum_{\beta\neq\alpha} \frac{1}{(v_{\alpha}-\varepsilon_i)(v_{\beta}-\varepsilon_j)} \frac{\det \mathbf{K}_{i\alpha,j\beta}}{\det\mathbf{K}} \\
P^{\tilde{u},v}_{ij} :=  \frac{\braket{\{\tilde{u}\} | S^+_i S^-_j | \{v\}}}{\braket{\{\tilde{u}\}|\{v\}}} 
&= \sum^{N_P}_{\alpha=1}\frac{1}{(v_{\alpha}-\varepsilon_j)} \left(
\frac{\det\mathbf{K}_{i\alpha}}{\det\mathbf{K}} - \sum_{\beta\neq\alpha} \frac{1}{v_{\beta}-\varepsilon_j} \frac{\det \mathbf{K}_{i\alpha,j\beta}}{\det\mathbf{K}}
\right)
\end{align}
If these formulas are implemented directly, then evaluating the 1RDM has computational cost $\mathcal{O}(N_PN^4_{orb})$ (to compute each nonzero element of the 1RDM one must evaluate $N_P$ determinants of $N_{orb}-1 \times N_{orb}-1$ matrices). The cost of constructing $D_{ij}$ and $P_{ij}$ is controlled by the cost of evaluating the double-sum: $\mathcal{O}(N^2_PN^5_{orb})$. (For each element of $D_{ij}$ and $P_{ij}$, one must evaluate $N^2_P$ determinants of $N_{orb}-2 \times N_{orb}-2$ matrices.) The actual cost is reduced because the orbitals, in methods like this, tend to be localized so $D_{ij}$ and $P_{ij}$ tend rapidly to zero when orbitals $i$ and $j$ are localized more than a few Angstroms away from each other.

The variational principle for the energy may be written as
\begin{align} \label{eq:OS_55}
E_{GS} &= \min_{ \substack {\{v\},\{\varepsilon\} \\ \ket{\{\tilde{u}\}} = \ket{\{v\}} }} 
\frac{\braket{\{\tilde{u}\}|\hat{H} |\{v\}}}{\braket{\{\tilde{u}\}|\{v\}}} \nonumber \\
&= \min_{ \substack {\{v\},\{\varepsilon\} \\ \ket{\{\tilde{u}\}} = \ket{\{v\}} }}
2\sum_i h_{ii} \gamma^{\tilde{u},v}_i + \sum_{i\neq j} (2V_{ijij}-V_{ijji})D^{\tilde{u},v}_{ij} + \sum_{ij} V_{iijj}  P^{\tilde{u},v}_{ij}
\end{align}
with the explicit requirement that the wavefunction be the same in its particle $\ket{\{\lambda\}}$ and hole $\ket{\{\tilde{\mu}\}}$ representations.

It should be clear from the preceding discussion that expressions for higher-order reduced
density matrices can be derived by the same approach with no additional difficulty. 

\subsection{Hyperbolic Off-Shell Bethe Vectors} \label{sec:OS_5}

The reduced BCS Hamiltonian, \eqref{eq:OS_22}, represents a completely isotropic interaction between
the pairs. There are more general Hamiltonians, with anisotropic interactions, that also have 
eigenvectors which may be determined using an ABA. For example, the XXZ Hamiltonian \cite{dunning,rombouts2,mario},
\begin{align} \label{eq:OS_56}
\hat{H}_{XXZ} = \frac{1}{2}\sum^{N_{orb}}_{i=1}\varepsilon_i \hat{n}_i  - \frac{g}{2}\sum^{N_{orb}}_{i,j=1}\sqrt{\varepsilon_i\varepsilon_j}S^+_iS^-_j
\end{align}
has eigenvectors of the geminal-product form,
\begin{align} \label{eq:OS_57}
\ket{\{u\},\{\eta\} } = \prod^{N_P}_{\alpha=1} S^+(u_a,\{\eta\})\ket{\theta}
\end{align}
where
\begin{align} \label{eq:OS_58}
S^+(u_a,\{\eta\}) = \sum^{N_{orb}}_{i=1} \frac{\eta_i}{u_a-\varepsilon_i}S^+_i.
\end{align}
In this particular case, $\eta_i=\sqrt{\varepsilon_i}$ and the rapidities $\{u\}$ satisfy the Bethe ansatz
equations, for each $a=1,\dots,N_P$
\begin{align} \label{eq:OS_59}
\frac{2}{g} + \sum^{N_{orb}}_{i=1}\frac{\varepsilon_i}{u_a-\varepsilon_i} + 2 \sum_{b \neq a}\frac{u_b}{u_b-u_a}=0.
\end{align}
It is not difficult to see that the solutions of Eqs. \eqref{eq:OS_25} and \eqref{eq:OS_59} must be distinct. While there are no additional parameters, the nature of the solutions is quite different and we are pursuing these models separately.

We consider here the off-shell case with arbitrary $\{\eta\}$. There are now
$2N_{orb} + N_P$ free parameters. Defining the dual state,
\begin{align} \label{eq:OS_60}
\bra{\{\tilde{u}\},\{\eta\}  } = \bra{\tilde{\theta}} \prod^{N_{orb}-N_P}_{a=1}S^+(\tilde{u}_a,\{\eta\}),
\end{align}
expressions for the scalar product and form factors can be worked out with the same techniques
employed in the previous section. The scalar product may be evaluated by writing it as the permanent
\begin{align}
\braket{ \{\tilde{u}\},\{\eta\} | \{v\},\{\eta\} } =
\begin{vmatrix}
\frac{\eta_1}{v_1 - \varepsilon_1} & \dots & \frac{\eta_{N_{orb}}}{v_1 - \varepsilon_{N_{orb}}} \\
\vdots & & \vdots \\
\frac{\eta_1}{v_{N_P} - \varepsilon_1} & \dots & \frac{\eta_{N_{orb}}}{v_{N_P} - \varepsilon_{N_{orb}}} \\
\frac{\eta_1}{\tilde{u}_1 - \varepsilon_1} & \dots & \frac{\eta_{N_{orb}}}{\tilde{u}_1 - \varepsilon_{N_{orb}}} \\
\vdots & & \vdots \\
\frac{\eta_1}{\tilde{u}_{N_{orb}-N_P} - \varepsilon_1} & \dots & \frac{\eta_{N_{orb}}}{\tilde{u}_{N_{orb}-N_P} - \varepsilon_{N_{orb}}}
\end{vmatrix}^+.
\end{align}
Since the permanent is linear in each column, the factor $\prod_k \eta_k$ can be removed, leaving the same permanent as the previous section. Therefore the scalar product is
\begin{align} \label{eq:hyp_scal}
\braket{ \{\tilde{u}\},\{\eta\} | \{v\},\{\eta\} } = \prod^{N_{orb}}_{k=1} \eta_k \det \textbf{K}.
\end{align}
The form factor approach is applied in the same manner, so that
\begin{align}
\frac{1}{2}\braket{ \{\tilde{u}\},\{\eta\} | \hat{n}_i | \{v\},\{\eta\} } &= \eta_i \sum_a \frac{ \braket{ \{\tilde{u}\},\{\eta\} | S^+_i | \{v\}_a,\{\eta\} } }{v_a - \varepsilon_i} \\
\frac{1}{4}\braket{ \{\tilde{u}\},\{\eta\} | \hat{n}_i \hat{n}_j | \{v\},\{\eta\} } &= \eta_i \eta_j \sum_{a\neq b} \frac{ \braket{ \{\tilde{u}\},\{\eta\} | S^+_i S^+_j | \{v\}_{a,b},\{\eta\} } }{(v_a - \varepsilon_i)(v_b-\varepsilon_j)} \\
\braket{ \{\tilde{u}\},\{\eta\} | S^+_i S^-_j | \{v\},\{\eta\} } &= \eta_j \sum_a \frac{ \braket{ \{\tilde{u}\},\{\eta\} | S^+_i | \{v\}_a,\{\eta\} } }{v_a - \varepsilon_j}
\nonumber \\
&- \eta_j \eta_j \sum_{a\neq b} \frac{ \braket{ \{\tilde{u}\},\{\eta\} | S^+_i S^+_j | \{v\}_{a,b},\{\eta\} } }{(v_a - \varepsilon_j)(v_b-\varepsilon_j)}.
\end{align}
The form factors are again calculated from the scalar product \eqref{eq:hyp_scal} using the solution to the inverse problem
\begin{align}
S^+_i = \frac{1}{\eta_i} \lim_{v_a \rightarrow \varepsilon_i} (v_a - \varepsilon_i)S^+(v_a,\{\eta\})
\end{align}
giving
\begin{align}
\braket{ \{\tilde{u}\},\{\eta\} | S^+_i | \{v\}_a,\{\eta\} } &= \prod_{k\neq i} \eta_k \det \textbf{K}_{ia} \\
\braket{ \{\tilde{u}\},\{\eta\} | S^+_i S^+_j | \{v\}_{a,b},\{\eta\} } &= \prod_{k\neq i,j} \eta_k \det \textbf{K}_{ia,jb}.
\end{align}
The normalized density matrix elements $\gamma^{\tilde{u},v,\eta}_i$ and $D^{\tilde{u},v,\eta}_{ij}$ end up identical to the rational off-shell versions, though the other elements are modified
\begin{align} \label{eq:OS_67}
P^{\tilde{u},v, \eta}_{ij} = \frac{\eta_j}{\eta_i}P^{\tilde{u},v}_{ij}.
\end{align}

One can generalize the wavefunction in Eqs. \eqref{eq:OS_57}-\eqref{eq:OS_58} by considering the most general form for which Borchardt's theorem holds,
\begin{align} \label{eq:OS_68}
\ket{\{v\},\{\eta\},\{\mu\}} = \prod^{N_P}_{a=1}\sum^{N_{orb}}_{i=1} \frac{\eta_i \mu_a}{v_a-\varepsilon_i}S^+_i\ket{\theta}.
\end{align}
However, this only changes the normalization of the wavefunction,
\begin{align} \label{eq:OS_69}
\ket{\{v\},\{\eta\},\{\mu\}} = \prod^{N_P}_{\beta=1}\mu_{\beta}
\prod^{N_P}_{\alpha=1}\sum^{N_{orb}}_{i=1} \frac{\eta_i }{v_{\alpha}-\varepsilon_i}S^+_i\ket{\theta}
= \prod^{N_P}_{\beta=1}\mu_{\beta} \ket{\{v\},\{\eta\}}
\end{align}
so none of the density matrix elements change.

\subsection{Nonorthogonal Orbitals} \label{sec:OS_6}

An alternative perspective on the hyperbolic model wavefunction, Eqs. \eqref{eq:OS_57}-\eqref{eq:OS_58}, is that
it relaxes the constraint that the orbitals be orthonormal. Specifically, the hyperbolic model is
equivalent to a rational model where the normalization of the orbitals is changed to $\eta_i$, but the
orbitals are still orthogonal. Can one generalize the rational model to general, nonorthogonal,
nonnormalized, single-particle states? Such a formulation is especially interesting because
electron-pair wavefunctions constructed from nonorthogonal orbitals are the fundamental
building block of elementary valence-bond calculations \cite{shaik2,wu,shaik1,gallup}.

We employ with spatial orbitals that are non-orthogonal, with second-quantized operators
\begin{align}
a^{\dagger}_{i\sigma} a_{j\tau} + a_{j\tau}a^{\dagger}_{i\sigma} = \Omega_{ij}\delta_{\sigma\tau}.
\end{align}
It is of course possible to perform the same construction with non-orthogonal spin-orbitals, though the notation becomes rather tedious quickly. The pair creators $S^+_i$ and annihilators $S^-_i$ no longer respect the su(2) structure. Moving an $S^+_i$ past an $S^-_j$ no longer produces $\delta_{ij}S^z_i$, but
\begin{align}
[S^+_i,S^-_j] = \Omega_{ij} \left( a^{\dagger}_{i\alpha} a_{j\alpha} + a^{\dagger}_{i\beta} a_{j\beta}  - \Omega_{ij}\right).
\end{align}
Scalar products and density matrices are still computable however. With the rational states $\ket{\{v\}}$ and $\bra{\{\tilde{u}\}}$ the scalar product becomes
\begin{align}
\braket{\{\tilde{u}\} | \{v\}} = \det \mathbf{\Omega} \det \mathbf{K}.
\end{align}
To evaluate the elements
\begin{align}
\gamma^{\tilde{u},v}_{ij} = \frac{\braket{\{\tilde{u}\} | \sum_{\sigma} a^{\dagger}_{i\sigma} a_{j\sigma} | \{v\}}}{\braket{\{\tilde{u}\} | \{v\}}}
\end{align}
the strategy is essentially the same. The numerator is evaluated by moving the one-body operator to the right until it destroys the vacuum. With the notation\cite{helgaker_book}
\begin{align}
\hat{E}_{ij} = \sum_{\sigma} a^{\dagger}_{i\sigma} a_{j\sigma}
\end{align}
\begin{align} \label{eq:no1_1dm}
\braket{\{\tilde{u}\} | \hat{E}_{ij} | \{v\} } = 
\sum_a \braket{\{\tilde{u}\} | [ \hat{E}_{ij} ,S^+(v_a) ] | \{v\}_a   },
\end{align}
where the commutator is easily evaluated
\begin{align}
[ \hat{E}_{ij} ,S^+(v_a) ] = \sum_m \frac{\Omega_{jm}}{(v_a - \varepsilon_m)}
(a^{\dagger}_{i\alpha}a^{\dagger}_{m\beta} + a^{\dagger}_{m\alpha}a^{\dagger}_{i\beta}).
\end{align}
Inserting this result in \eqref{eq:no1_1dm}, only the terms for which $i=m$ survive as all others will attempt to create an electron more than once. This leads to the result
\begin{align}
\gamma^{\tilde{u},v}_{ij} = 2 \frac{\Omega_{ji}}{\det \mathbf{\Omega}} \sum_a \frac{1}{(v_a - \varepsilon_i)} \frac{\det \mathbf{K}_{ia}}{\det \mathbf{K}}.
\end{align}
The elements of the 2-RDM are evaluated similarly. Denoting,
\begin{align}
\hat{e}_{ijkl} = \sum_{\sigma\tau} a^{\dagger}_{i\sigma}a^{\dagger}_{j\tau}a_{l\tau}a_{k\sigma}
\end{align}
the elements
\begin{align}
\Gamma^{\tilde{u},v}_{ijkl} = \frac{\braket{\{\tilde{u}\} | \hat{e}_{ijkl} |\{v\} }}{\braket{\{\tilde{u}\}|\{v\} }}
\end{align}
are evaluated by moving $\hat{e}_{ijkl}$ to the right
\begin{align}
\braket{ \{\tilde{u}\}| \hat{e}_{ijkl} | \{v\}} &= \sum_a \braket{\{\tilde{u}\} | \prod_{b\neq a} S^+(v_b) [\hat{e}_{ijkl},S^+(v_a)] | \theta } \nonumber \\
&+ \sum_{a < b} \braket{ \{\tilde{u}\} | [[\hat{e}_{ijkl},S^+(v_a)],S^+(v_b)] | \{v\}_{a,b} }.
\end{align}
The action of the single commutator on the vacuum is
\begin{align}
[\hat{e}_{ijkl},S^+(v_a)] \ket{\theta} = \sum_m \frac{\Omega_{km}\Omega_{lm}}{(v_a - \varepsilon_m)} (a^{\dagger}_{i\alpha}a^{\dagger}_{j\beta} + a^{\dagger}_{j\alpha}a^{\dagger}_{i\beta})\ket{\theta},
\end{align}
and the double commutator is
\begin{align}
[[\hat{e}_{ijkl},S^+(v_a)],S^+(v_b)] &= 
\sum_{mn} \frac{\Omega_{km}\Omega_{ln} - \Omega_{kn}\Omega_{lm} }{(v_a-\varepsilon_m)(v_b - \varepsilon_n)}
(a^{\dagger}_{i\alpha}a^{\dagger}_{m\beta}a^{\dagger}_{j\alpha}a^{\dagger}_{n\beta}+ a^{\dagger}_{m\alpha}a^{\dagger}_{i\beta}a^{\dagger}_{n\alpha}a^{\dagger}_{j\beta}) \nonumber \\
&+ \sum_{mn} \frac{\Omega_{km}\Omega_{ln}}{(v_a - \varepsilon_m)(v_b - \varepsilon_n)}
(a^{\dagger}_{i\alpha}a^{\dagger}_{m\beta}a^{\dagger}_{n\alpha}a^{\dagger}_{j\beta} + a^{\dagger}_{m\alpha}a^{\dagger}_{i\beta}a^{\dagger}_{j\alpha}a^{\dagger}_{n\beta} ) \nonumber \\
&+ \sum_{mn} \frac{\Omega_{kn}\Omega_{lm}}{(v_a - \varepsilon_m)(v_b - \varepsilon_n)}
(a^{\dagger}_{i\alpha}a^{\dagger}_{n\beta}a^{\dagger}_{m\alpha}a^{\dagger}_{j\beta} + a^{\dagger}_{n\alpha}a^{\dagger}_{i\beta}a^{\dagger}_{j\alpha}a^{\dagger}_{m\beta} ).
\end{align}
Now, the $i=j$ and $i\neq j$ cases must be treated separately. For $i=j$, the single commutator acting on the vacuum contributes as well as the second and third summations from the double commutator. The result is
\begin{align}
\Gamma^{\tilde{u},v}_{iikl} &= 2\sum_a \sum_m \frac{\Omega_{km}\Omega_{lm}}{\det \mathbf{\Omega}} \frac{1}{(v_a-\varepsilon_m )} \frac{\det \mathbf{K}_{ia}}{\det \mathbf{K}} \nonumber \\
&-2 \sum_m \sum_{a \neq b} \frac{\Omega_{km}\Omega_{lm}}{\det \mathbf{\Omega}} \frac{1}{(v_a - \varepsilon_m)(v_b - \varepsilon_m)} \frac{\det \mathbf{K}_{iamb}}{\det \mathbf{K}},
\end{align}
where the second set of terms arise from the double commutators, with the condition that $m=n$. For $i\neq j$, only the double commutator terms contribute. In the first summation, there are two contributions, from $i=m,\;j=n$ and $i=n,\;j=m$, while the other summations give unique contributions. The final result, after rearranging some summations, is
\begin{align}
\Gamma^{\tilde{u},v}_{ijkl} &= 2 \sum_{a \neq b} \frac{(2\Omega_{ki}\Omega_{lj}-\Omega_{li}\Omega_{kj})}{\det \mathbf{\Omega}} \frac{1}{(v_a-\varepsilon_i)(v_b-\varepsilon_j)}
\frac{\det \mathbf{K}_{iajb}}{\det \mathbf{K}}
\end{align}
where in particular, we've used
\begin{align}
\sum_{a < b} \left( \frac{1}{(v_a-\varepsilon_i)(v_b-\varepsilon_j)} + \frac{1}{(v_a-\varepsilon_j)(v_b-\varepsilon_i)} \right)
= \sum_{a \neq b} \frac{1}{(v_a-\varepsilon_i)(v_b-\varepsilon_j)}.
\end{align}
These results are asymmetric as we have privileged the set $\{v\}$ in their derivation.

\section{Numerical Considerations} \label{sec:OS_7}
\subsection{Preliminaries} \label{sec:OS_7p1}

The results from sections \ref{sec:OS_2}-\ref{sec:OS_4} indicate that one may variationally optimize the off-shell
Bethe vectors associated with the Richardson-Gaudin structure provided that two different representations of the state are simultaneously known. In order to give explicit procedures
for doing this, it is useful to write the rapidities of the dual state as
\begin{align} \label{eq:OS_74}
v_{N_P+a} = \tilde{u}_a
\end{align}
With this notational convention, the wavefunction and its dual are written as
\begin{align} \label{eq:OS_75}
\ket{\{v_{\alpha}\}^{N_P}_{\alpha=1}, \{\varepsilon_i,\eta_i\}^{N_{orb}}_{i=1}}
= \prod^{N_P}_{\alpha=1}\sum^{N_{orb}}_{i=1} \frac{\eta_i}{v_{\alpha}-\varepsilon_i}S^+_i\ket{\theta}
\end{align}
and
\begin{align} \label{eq:OS_76}
\ket{\{v_{\alpha}\}^{N_{orb}}_{\alpha=N_P+1}, \{\varepsilon_i,\eta_i\}^{N_{orb}}_{i=1}}
= \prod^{N_{orb}}_{\alpha=N_P+1}\sum^{N_{orb}}_{i=1} \frac{\eta^*_i}{v^*_{\alpha}-\varepsilon_i}S^-_i\ket{\tilde{\theta}},
\end{align}
respectively. 

\subsection{Bivariational Principle} \label{sec:OS_7p2}

The bivariational principle was introduced to quantum chemistry by Boys and Handy \cite{boys},
and is relatively well-known in (albeit rarely used by) the coupled-cluster community \cite{bernardi,kutz3,arponen,kvaal2,kvaal1,basughose,pal}. It
indicates that the asymmetric energy expectation value expression,
\begin{align} \label{eq:OS_77}
E[\Psi,\tilde{\Psi}] = \frac{\braket{\tilde{\Psi}|\hat{H}|\Psi}}{\braket{\tilde{\Psi}|\Psi}}
\end{align}
is stationary with respect to variations of both $\ket{\Psi}$ and $\ket{\tilde{\Psi}}$
\begin{align} \label{eq:OS_78}
\frac{\delta E[\Psi,\tilde{\Psi}]}{\delta \Psi} = \frac{\delta E[\Psi,\tilde{\Psi}]}{\delta \tilde{\Psi}}=0
\end{align}
only when $\ket{\Psi}$ and $\ket{\tilde{\Psi}}$ are equal to each other (i.e., they are dual) and they are equal to an
eigenstate of the Hamiltonian. (Matters are slightly more complicated for degenerate states.)
Applying this principle to our off-shell RG states using the notation
from the previous section, one needs to solve $3N_{orb}$ nonlinear equations in $3N_{orb}$ unknowns,
\begin{align} \label{eq:OS_79}
\frac{\partial E[\{ \lambda_i,\eta_i,\varepsilon_i \}^{N_{orb}}_{i=1}]}{\partial \lambda_i}=0 \nonumber \\
\frac{\partial E[\{ \lambda_i,\eta_i,\varepsilon_i \}^{N_{orb}}_{i=1}]}{\partial \eta_i}=0  \\
\frac{\partial E[\{ \lambda_i,\eta_i,\varepsilon_i \}^{N_{orb}}_{i=1}]}{\partial \varepsilon_i}=0 \nonumber 
\end{align}
In practical calculations, one would also optimize over the choice of orbitals. This could be done
together with the optimization of the wavefunction parameters or alternately, iteratively
optimizing the wavefunction parameters (Eq. \eqref{eq:OS_79}) and then minimizing the resulting energy
expression with respect to the orbitals until convergence \cite{stein,kasia_OO}.

The bivariational principle does not provide a lower bound to the true energy; the energy
expression in Eq. \eqref{eq:OS_77} is not bounded from below. It is not even guaranteed that the stationary
values for the energy are real (though this could be added as a variational constraint). If,
however, the solution of Eqs. \eqref{eq:OS_79} gives wavefunctions that are dual,
\begin{align} \label{eq:OS_80}
\prod^{N_P}_{\alpha=1}\sum^{N_{orb}}_{i=1} \frac{\eta_i}{v_{\alpha}-\varepsilon_i}S^+_i\ket{\theta} =
\prod^{N_{orb}}_{\alpha=N_P+1}\sum^{N_{orb}}_{i=1} \frac{\eta^*_i}{v^*_{\alpha}-\varepsilon_i}S^-_i\ket{\tilde{\theta}},
\end{align}
then the stationary-value for the energy is an upper bound to the true ground-state energy.

\subsection{Dual-Constrained Minimization} \label{sec:OS_7p3}

The failsafe method for variationally optimizing the off-shell RG wavefunction is to minimize
the energy expression subject to the duality constraint, Eq. \eqref{eq:OS_80}. (See, for example,
Eq. \eqref{eq:OS_55}.) We have been unable to find any practical way to exactly enforce the duality
constraint. Notice, however, that the duality constraint is valid if and only if
\begin{align} \label{eq:OS_81}
\Braket{\Phi|\prod^{N_P}_{\alpha=1}\sum^{N_{orb}}_{i=1} \frac{\eta_i}{v_{\alpha}-\varepsilon_i}S^+_i | \theta} =
\Braket{\Phi|\prod^{N_{orb}}_{\alpha=N_P+1}\sum^{N_{orb}}_{i=1} \frac{\eta^*_i}{v^*_{\alpha}-\varepsilon_i}S^-_i|\tilde{\theta}} \;\;\;\forall\Phi
\end{align}
Equation \eqref{eq:OS_81} can be easily evaluated if $\Phi$ is a Slater determinant. Specifically, both sides of the
equation are zero if any electron in the Slater determinant is unpaired. If the Slater determinant
has the same pairing structure as the off-shell RG state, then Eq. \eqref{eq:OS_81} is an identity about the
permanents of the coefficient matrices,
\begin{align} \label{eq:OS_82}
\vert\mathbf{C}_{\Phi}\vert^+ = \vert\tilde{\mathbf{C}}_{\Phi}\vert^+
\end{align}
where the $N_P \times N_P$ matrix $\mathbf{C}_{\Phi}$ includes orbitals that are occupied in $\Phi$ and the $(N_{orb} - N_P) \times (N_{orb} - N_P)$ matrix $\tilde{\mathbf{C}}_{\Phi}$ includes orbitals that are not occupied in $\Phi$. I.e.,
\begin{align} \label{eq:OS_83}
\mathbf{C}_{\Phi} =
\left[\begin{array}{cccc}
\frac{\eta_{i_1\in\Phi}}{v_1 - \varepsilon_{i_1\in\Phi}} & \frac{\eta_{i_2\in\Phi}}{v_1 - \varepsilon_{i_2\in\Phi}} &
\dots & \frac{\eta_{i_{N_P}\in\Phi}}{v_1 - \varepsilon_{i_{N_P}\in\Phi}} \\
\frac{\eta_{i_1\in\Phi}}{v_2 - \varepsilon_{i_1\in\Phi}} & \frac{\eta_{i_2\in\Phi}}{v_2 - \varepsilon_{i_2\in\Phi}} &
\dots & \frac{\eta_{i_{N_P}\in\Phi}}{v_2 - \varepsilon_{i_{N_P}\in\Phi}} \\
\vdots & \vdots & \ddots & \vdots \\
\frac{\eta_{i_1\in\Phi}}{v_{N_P} - \varepsilon_{i_1\in\Phi}} & \frac{\eta_{i_2\in\Phi}}{v_{N_P} - \varepsilon_{i_2\in\Phi}} & \dots & \frac{\eta_{i_{N_P}\in\Phi}}{v_{N_P} - \varepsilon_{i_{N_P}\in\Phi}} \\
\end{array} \right]
\end{align}
\begin{align} \label{eq:OS_84}
\tilde{\mathbf{C}}_{\Phi} =
\left[\begin{array}{cccc}
\frac{\eta^*_{i_1\notin\Phi}}{v^*_{N_P+1} - \varepsilon_{i_1\notin\Phi}} & 
\frac{\eta^*_{i_2\notin\Phi}}{v^*_{N_P+1} - \varepsilon_{i_2\notin\Phi}} & \dots &
\frac{\eta^*_{i_{N_{orb}-N_P}\notin\Phi}}{v^*_{N_P+1} - \varepsilon_{i_{N_{orb}-N_P}\notin\Phi}} \\
\frac{\eta^*_{i_1\notin\Phi}}{v^*_{N_P+2} - \varepsilon_{i_1\notin\Phi}} & 
\frac{\eta^*_{i_2\notin\Phi}}{v^*_{N_P+2} - \varepsilon_{i_2\notin\Phi}} & \dots &
\frac{\eta^*_{i_{N_{orb}-N_P}\notin\Phi}}{v^*_{N_P+2} - \varepsilon_{i_{N_{orb}-N_P}\notin\Phi}} \\
\vdots & \vdots & \ddots & \vdots \\
\frac{\eta^*_{i_1\notin\Phi}}{v^*_{N_{orb}} - \varepsilon_{i_1\notin\Phi}} & 
\frac{\eta^*_{i_2\notin\Phi}}{v^*_{N_{orb}} - \varepsilon_{i_2\notin\Phi}} & \dots &
\frac{\eta^*_{i_{N_{orb}-N_P}\notin\Phi}}{v^*_{N_{orb}} - \varepsilon_{i_{N_{orb}-N_P}\notin\Phi}}
\end{array} \right]
\end{align}
The permanents of these rank-2 matrices can be efficiently evaluated. Specifically, Eq. \eqref{eq:OS_82}
reduces to the constraint,
\begin{align} \label{eq:OS_85}
\mathfrak{d}[\Phi] = \frac{\vert \mathbf{C}_{\Phi} * \mathbf{C}_{\Phi} \vert}{\vert \mathbf{C}_{\Phi}\vert}-
\frac{\vert \tilde{\mathbf{C}}_{\Phi} * \tilde{\mathbf{C}}_{\Phi} \vert}{\vert \tilde{\mathbf{C}}_{\Phi}\vert}=0
\end{align}
where $*$ again denotes the Hadamard (element-wise) product.

It is not practical to impose the duality constraint in Eq. \eqref{eq:OS_85} for all Slater determinants.
In the spirit of our work using the projected Schr\"{o}dinger equation to optimize the
parameters in the off-shell RG state, we only require Eq. \eqref{eq:OS_85} to hold for a reference
determinant

\begin{align} \label{eq:OS_86}
\ket{\Phi_0} = \ket{1\bar{1}2\bar{2}\dots N_P\bar{N}_P}
\end{align}
and its one-pair excitations,
\begin{align} \label{eq:OS_87}
\ket{\Phi^{a\bar{a}}_{i\bar{i}}} = S^+_aS^-_i \ket{1\bar{1}2\bar{2}\dots N_P\bar{N}_P},\;\;\;1\leq i\leq N_P \leq a \leq N_{orb}.
\end{align}
In comparison to our previous work, we are using an approximate expression for duality instead
of approximately forcing the wavefunction to satisfy the Schr\"{o}dinger equation. Even with this
restriction, Eq. \eqref{eq:OS_85} represents $N_P(N_{orb} - N_P)$ equations with only $3N_P$ unknowns; we believe that
it is very unlikely to find a solution to this equation that are not true dual vectors. If one is
suspicious that spurious false-dual vectors have been found, one can add consider pair-excitations
from additional reference Slater determinants.

Introducing Lagrange multipliers for the duality constraints, the duality constrained
variational principle corresponds to optimizing the Lagrangian
\begin{align} \label{eq:OS_88}
\mathcal{L}[\{v,\varepsilon,\eta\};\{\xi\} ] = E[\{v,\varepsilon,\eta\}]+ \xi_0 \mathfrak{d}[\Phi_0]
+ \sum^{N_P}_{i=1}\sum^{N_{orb}}_{a=N_P+1} \xi^a_i \mathfrak{d}[\Phi^{a\bar{a}}_{i\bar{i}}].
\end{align}
Even though the equations we use to force duality are overdetermined, solutions always
exist because every eigenstate of a Richardson-like Hamiltonian, Eqs. \eqref{eq:OS_22} or \eqref{eq:OS_56}, has a dual
representation. Specifically, given an eigenvector of Richardson's equations, $\ket{\{v\}}$, the dual
vector is the solution of the overdetermined set of nonlinear equations \cite{faribault_arx},
\begin{align} \label{eq:OS_89}
\sum^{N_{orb}}_{a=N_P+1}\frac{\eta_i}{\varepsilon_i - v_a}=
\sum^{N_P}_{a=1}\frac{\eta_i}{\varepsilon_i - v_a} -\frac{2}{g}, \quad \forall i.
\end{align}
It is unclear whether Eqs. \eqref{eq:OS_89} are necessary and sufficient for duality. Given an arbitrary off-shell
eigenvector, however, one could construct a near-dual state by solving the equations,
\begin{align} \label{eq:OS_90}
\sum^{N_P}_{a=1}\frac{\eta_i}{\varepsilon_i - v_a} - \sum^{N_{orb}}_{a=N_P+1}\frac{\eta_i}{\varepsilon_i - v_a} =\bar{c}
\end{align}
where
\begin{align} \label{eq:OS_91}
\bar{c} = \frac{1}{N_{orb}} \sum^{N_{orb}}_{i=1}
\left(\sum^{N_P}_{a=1}\frac{\eta_i}{\varepsilon_i - v_a} - \sum^{N_{orb}}_{a=N_P+1}\frac{\eta_i}{\varepsilon_i - v_a}
\right).
\end{align}

\subsection{Projected Schr\"{o}dinger Equation} \label{sec:OS_7p4}

These approaches can be contrasted with the approach we have used previously, based on
the projected Schr\"{o}dinger equation \cite{johnson,limacher,kasia_OO,kasia_hubbard,pawel_geminal,stein}. In the projected Schr\"{o}dinger equation, we examine the weak
formulation of the eigenvalue problem, namely,
\begin{align} \label{eq:OS_92}
\braket{\tilde{\Psi}|\hat{H}|\{v,\varepsilon,\eta\}} = E\braket{\tilde{\Psi}|\{v,\varepsilon,\eta\}},\quad\forall \tilde{\Psi}.
\end{align}
That is, one projects the Schr\"{o}dinger equation against all possible wavefunctions, and forces
the expected value of the energy to be the same in all cases.

In our previous work, we projected wavefunctions with the form \eqref{eq:OS_13} against Slater
determinants \cite{johnson,limacher,kasia_OO,kasia_hubbard,pawel_geminal,stein}. Since the action of the Hamiltonian on a Slater determinant is just a linear
combination of Slater determinants, this procedure had computational cost $\mathcal{O}(N^3_P(N_{orb}-N_P)N_{orb})$. Now it is clear, however, that we could also project against off-shell XXZ wavefunctions,
\begin{align} \label{eq:OS_93}
\braket{\{\tilde{u},\varepsilon,\eta\}|\hat{H}|\{v,\varepsilon,\eta\}} = E
\braket{\{\tilde{u},\varepsilon,\eta\}|\{v,\varepsilon,\eta\}}
\end{align}
Evaluating each of these nonlinear equations is just as difficult as evaluating the energy itself,
and one needs at least one equation for each unknown parameter in the wavefunction. (Often one
would choose more equations than unknowns, and then determine the least-squares solution to
the overdetermined system of equations.) The cost of this procedure is at least $\mathcal{O}(N^2_PN^5_{orb})$, but it
will be more robust than projection against Slater determinants when the orbital picture breaks
down so completely that it is difficult to pick a suitable set of Slater determinants to project
against.

\section{Special Case: The Antisymmetrized Geminal Power} \label{sec:OS_8}

The antisymmetrized geminal power (AGP) wavefunction arises when all the geminals
are equal \cite{coleman1}. Referring to the wavefunction form \eqref{eq:OS_57}-\eqref{eq:OS_58}, we must set all the rapidities equal to
each other. However, it is convenient to retain both $\{\varepsilon\}$ and $\{\eta\}$ as (redundant) free parameters,
\begin{align} \label{eq:OS_94}
\ket{\text{AGP}} = \left( \sum^{N_{orb}}_{i=1}\frac{\eta_i S^+_i}{v-\varepsilon_i} \right)^{N_P}\ket{\theta}
\equiv \left( \sum^{N_{orb}}_{i=1} c_iS^+_i \right)^{N_P}\ket{\theta}
\end{align}
To obtain an explicit energy expression, we need to find the dual representation of this
state. We will demonstrate, later, that the dual representation is also an AGP, i.e.,
\begin{align} \label{eq:OS_95}
\ket{\text{AGP}} = \left( \sum^{N_{orb}}_{i=1}\frac{\eta^*_i S^-_i}{u-\varepsilon_i} \right)^{N_{orb}-N_P}\ket{\tilde{\theta}}
\equiv \left( \sum^{N_{orb}}_{i=1} \tilde{c}_iS^-_i \right)^{N_{orb}-N_P}\ket{\tilde{\theta}}.
\end{align}
The two representations of the wavefunction are equivalent if, for any Slater determinant in
which all electrons are paired, the following equation holds:
\begin{align} \label{eq:OS_96}
\Braket{\Phi|\left( \sum^{N_{orb}}_{i=1} c_iS^+_i \right)^{N_P} |\theta} =
\Braket{\Phi|\left( \sum^{N_{orb}}_{i=1} \tilde{c}_iS^-_i \right)^{N_{orb}-N_P} |\tilde{\theta}}.
\end{align}
Referring to section \ref{sec:OS_7p2}, this implies an identity about the permanents of the coefficient
matrices, with the simplifying feature that every row in the matrices \eqref{eq:OS_83} and \eqref{eq:OS_84} is the same.
Using the formula for the permanent of a rank-one matrix, Eq. \eqref{eq:OS_96} can be rewritten as
\begin{align} \label{eq:OS_97}
N_P!\prod_{i\in\Phi}c_i=(N_{orb}-N_P)!\prod_{i\notin\Phi}\tilde{c}_i.
\end{align}
The solution to this equation is
\begin{align} \label{eq:OS_98}
\tilde{c}_i = \frac{\left( \frac{N_P!}{(N_{orb}-N_P)!}\prod^{N_{orb}}_{k=1}c_k \right)^{\frac{1}{N_{orb}-N_P}}}{c_i} 
= \frac{\mathcal{K}}{c_i}
\end{align}
as may be verified by direct substitution into Eq. \eqref{eq:OS_97}. Note that the existence of this explicit
solution confirms that the dual representation of an AGP, Eq. \eqref{eq:OS_94}, is also an AGP, Eq. \eqref{eq:OS_95}.
The numerator of Eq. \eqref{eq:OS_98} is a constant, $\mathcal{K}$, which is arbitrary since choosing its value is
controlled by the phase and normalization of the wavefunction. 
Finding the dual representation in the XXZ form of the AGP is equivalent to solving two equations and two
unknowns. For convenience, fix the scale and zero of energy by setting $\lambda=0$ and $\mu=1$. Then,
from Eqs. \eqref{eq:OS_94}, \eqref{eq:OS_95}, and \eqref{eq:OS_98},
\begin{align} \label{eq:OS_100}
c_i &= -\frac{\eta_i}{\varepsilon_i} \nonumber \\
c_i\tilde{c}_i &= \frac{\vert\eta\vert^2}{\varepsilon_i(1-\varepsilon^*_i)}=\mathcal{K}
\end{align}
which have the solutions
\begin{align} \label{eq:OS_101}
\varepsilon_i &= \frac{\mathcal{K}^*}{\vert c_i\vert^2 +\mathcal{K}^*} \nonumber \\
\eta_i &= -\frac{\mathcal{K}^*c_i}{\vert c_i\vert^2 +\mathcal{K}^*}.
\end{align}
This gives an explicit expression for the energy that can be minimized as a function of $c_i$. It is
simpler, on paper, to derive expressions if one chooses $\{\varepsilon_i\}$ to be real-valued variational
parameters, imposes the duality condition
\begin{align} \label{eq:OS_102}
\eta_i = \sqrt{\varepsilon_i(1-\varepsilon_i)},
\end{align}
and minimizes the energy as a function of $\{\varepsilon_i\}$.

It bears mentioning that the AGP can be regarded as an eigenvector of an XXZ Hamiltonian at the Moore-Read point, where there is a condensation of pair energies \cite{dunning,rombouts2,mario}. This is worked out explicitly in the appendix of ref. \cite{mario}.

\section{Discussion} \label{sec:OS_9}

In the theory of electronic structure, there are very few wavefunction-families
that lend themselves to variational optimization. We believe that an ideal variational method
should be size-consistent and that its computational requirements should grow only as a
(preferably small) polynomial in the size of the system. While there is some debate about
whether size-consistent variational methods exist \cite{hirata}, we believe that the approaches presented
here meet these requirements, as on-shell RG states do.\cite{fecteau:2022} Not many previous methods do, and one could argue that aside
from full-CI, the only variational size-consistent methods are mean-field models for
(quasi)particles, like the methods considered in this paper. Other polynomial-scaling and
variational post-Hartree-Fock methods (e.g., limited configuration interaction) tend not to be size-consistent. Polynomial and size-consistent post Hartree-Fock methods (e.g., coupled cluster
methods, many body perturbation theory) tend not to be variational. The density matrix
renormalization group (DMRG) algorithm for optimizing matrix product states is variational and
size-consistent, but has exponentially-growing computational cost for most three-dimensional
molecular systems \cite{garnet2,white1,white2,garnet1}.

Our approach fits best into the hierarchy of geminal-based approaches. The
antisymmetric product of strongly orthogonal geminals wavefunction is also a size-consistent,
variational, and polynomial-scaling method. The antisymmetric
geminal power wavefunction is variational and polynomial-scaling, but not size-consistent. The question arises: can we stretch the geminals formulation while retaining (a) a polynomial-cost
variational method, (b) size-consistency, and (c) a theory that includes all of the more traditional
geminal-based wavefunctions as special cases? The methods presented here, based on the
variational optimization of off-shell Cauchy geminals, do exactly this.

To what extent can the results in this paper be generalized even further? For example, in
reference \cite{johnson}, we used the superalgebras, gl(m$|$n), to develop approaches for open-shell atoms and
molecules. The treatment presented in this paper can be extended to these superalgebras, but the
associated method has factorial computational scaling. We have explored many other algebras
also, but factorial computational cost seems to arise for all algebras except su(2).

Can we variationally optimize more general wavefunctions based on su(2)? For example,
in reference \cite{limacher}, we used the projected Schr\"{o}dinger equation to explore an su(2)-based
wavefunction with the form,
\begin{align} \label{eq:OS_103}
\ket{\text{AP1roG}} = \prod^{N_P}_{a=1}\left( \delta_{ai}+\sum^{N_{orb}}_{i=N_P+1}g_{a,i} \right)S^+_i\ket{\theta}
\end{align}
This wavefunction, which we call AP1roG (antisymmetric product of 1-reference-orbital
geminals) has proved to be highly effective for many systems, including some that are strongly
correlated. Can we use the methods in this paper to formulate a variational version of that
theory?

A variational version of AP1roG indeed exists. It even has the desirable property that the
dual vector is easily constructed. Specifically, the dual vectors are,
\begin{align} \label{eq:OS_104}
\bra{\text{AP1roG}} = \bra{\tilde{\theta}}\prod^{N_{orb}}_{\alpha=N_P+1}\left(\delta_{i\alpha}+\sum^{N_P}_{i=1}g^*_{i,\alpha} \right)S^-_i
\end{align}
Unfortunately, evaluating permanents of matrices with the form
\begin{align} \label{eq:OS_105}
\mathbf{G}_{\text{AP1roG}} = \left[\begin{array}{cc}
\mathbf{I}_{N_P\times N_P} & \mathbf{G}_{N_P\times (N_{orb}- N_P)} \\
\mathbf{G}^{\dagger}_{(N_{orb}- N_P)\times N_P } & \mathbf{I}_{(N_{orb}- N_P)\times (N_{orb}- N_P)}
\end{array} \right]
\end{align}
is computational intractable, so variational optimization of the AP1roG wavefunction has
factorial scaling. It has however shown promise when targeting states specifically.\cite{marie:2021,kossoski:2021}

\section{Conclusion} \label{sec:OS_10}
We have developed (bi-)variational approaches for geminal product wavefunctions from the algebraic structure of su(2).  After considering the case where the wavefunction is an eigenfunction of Richardson-Gaudin type Hamiltonians (on-shell; section \ref{sec:OS_3}), we note that the reduced density matrices between wavefunctions that have the Richardson-Gaudin wavefunction form can also be evaluated (off-shell, section \ref{sec:OS_4}). We then generalize this wavefunction form to unnormalized (hyperbolic case, section \ref{sec:OS_5}) and nonorthogonal (section \ref{sec:OS_6}) orbitals. All of these methods are based on the same key ``tricks:'' elements of the reduced density matrices are written as expectation values of su(2) operators. These expectation values are then rewritten as projections onto the fully-filled Slater determinant. This trick allows us to exploit the only cheap result that we know for a general su(2) wavefunction: the projection of an su(2) wavefunction onto a Slater determinant is a permanent (cf. Eq. \eqref{eq:OS_14}), which is easy to evaluate with the Cauchy structure. In general, scalar products and density matrix elements of su(2) wavefunctions are much more complicated.\cite{moisset:2022}

Expressions for the density matrices are used to propose bivariational methods for determining the energy. The alternative, and more rigorous, approach is to minimize the energy subject to the constraint that the wavefunction expression in the ket (which generates electrons from the vacuum) and the wavefunction expression in the bra (which generates holes in the fully-filled Slater determinant) are equal. Both of these (bi-)variational approaches have higher computational scaling than previously proposed methods based on the projected Schr\"{o}dinger equation. As an application of this result, we developed a new, direct, and explicit, variational optimization procedure for the antisymmetrized geminal power (AGP) wavefunction.

\section{Acknowledgements}
P.A.J. was supported by NSERC and Compute Canada. P.W.A. acknowledges funding from NSERC, Compute Canada and the Canada Research Chairs. S.D.B. thanks the Canada Research Chairs.

\appendix
\section{Energy Formulas}
This appendix summarizes the key formulas for the energy. Using orbitals that are orthonormal, the energy, expressed as a function of the wavefunction parameters is
\begin{align}
E[\{v_i, \varepsilon_i,\eta_i\}^{N_{orb}}_{i=1}] &= \frac{\braket{\{v_a\}^{N_{orb}}_{a=1},\{\varepsilon_i,\eta_i\}^{N_{orb}}_{i=1}| \hat{H}_C | \{v_a\}^{N_P}_{a=1},\{\varepsilon_i,\eta_i\}^{N_{orb}}_{i=1} }}{\braket{\{v_a\}^{N_{orb}}_{a=1},\{\varepsilon_i,\eta_i\}^{N_{orb}}_{i=1} | \{v_a\}^{N_P}_{a=1},\{\varepsilon_i,\eta_i\}^{N_{orb}}_{i=1} }} \\
&= 2 \sum^{N_{orb}}_{i=1} h_{ii} \gamma_{i} + \sum_{i\neq j} (2V_{ijij}-V_{ijji}) D_{ij} + \sum_{ij} V_{iijj} P_{ij}
\end{align}
where the 1-density matrix $\gamma_{i}$ is diagonal
\begin{align}
\gamma_{i} = \sum^{N_P}_{a=1} \frac{1}{v_a - \varepsilon_i} \frac{\det \mathbf{K}_{ia}}{\det \mathbf{K}},
\end{align}
and for $i\neq j$,
\begin{align}
D_{ij} &= \sum_{a \neq b} \frac{1}{(v_a - \varepsilon_i)(v_b - \varepsilon_j)} \frac{\det \mathbf{K}_{ia,jb}}{\det \mathbf{K}} \\
P_{ij} &= \frac{\eta_j}{\eta_i} \left(
\sum^{N_P}_{a=1} \frac{1}{v_a - \varepsilon_j} \frac{\det \mathbf{K}_{ia}}{\det \mathbf{K}}-\sum_{a \neq b} \frac{1}{(v_a-\varepsilon_j)(v_b-\varepsilon_j)}
\frac{\det \mathbf{K}_{ia,jb}}{\det \mathbf{K}}.
\right)
\end{align}
The diagonal elements $D_{ii}$ and $P_{ii}$ refer to the same matrix element, so it is assigned to $P_{ii}=\gamma_{i}$. The elements of $\mathbf{K},\mathbf{K}_{ia}$ and $\mathbf{K}_{ia,jb}$ are
\begin{align}
k_{\alpha\beta} = \begin{cases}
\sum^{N_{orb}}_{\substack{k=1 \\ k\neq \alpha}} \frac{1}{\varepsilon_{\alpha}-\varepsilon_k}- \sum^{N_{orb}}_{a=1} \frac{1}{\varepsilon_{\alpha}-v_a}, \quad & \alpha = \beta \\
\frac{1}{\varepsilon_{\alpha} - \varepsilon_{\beta}}, \quad & \alpha \neq \beta
\end{cases}
\end{align}

\begin{align}
[k_{ia}]_{\alpha\beta} = \begin{cases}
\sum^{N_{orb}}_{\substack{k=1 \\ k\neq i,\alpha}} \frac{1}{\varepsilon_{\alpha}-\varepsilon_k}- \sum^{N_{orb}}_{\substack{c=1\\c\neq a}} \frac{1}{\varepsilon_{\alpha}-v_a}, \quad & \alpha = \beta (\neq i) \\
\frac{1}{\varepsilon_{\alpha} - \varepsilon_{\beta}}, \quad & \alpha \neq \beta (\neq i)
\end{cases}
\end{align}

\begin{align}
[k_{ia,jb}]_{\alpha\beta} = \begin{cases}
\sum^{N_{orb}}_{\substack{k=1 \\ k\neq i,j,\alpha}} \frac{1}{\varepsilon_{\alpha}-\varepsilon_k}- \sum^{N_{orb}}_{\substack{c=1 \\ c \neq a,b}} \frac{1}{\varepsilon_{\alpha}-v_c}, \quad & \alpha = \beta (\neq i,j) \\
\frac{1}{\varepsilon_{\alpha} - \varepsilon_{\beta}}, \quad & \alpha \neq \beta (\neq i,j)
\end{cases}
\end{align}
The gradient of the energy with respect to the parameters in the wavefunction may be evaluated in a straightforward manner.

\bibliographystyle{unsrt}
\bibliography{off_shell}

\end{document}